\documentclass[11pt]{article}

\usepackage{graphics} 
\usepackage{epsfig} 
\usepackage{times} 
\usepackage{amsmath} 
\usepackage{amsthm}
\usepackage{amssymb}
\usepackage{tikz}
\usetikzlibrary{graphs}
\usepackage[colorlinks=false, urlcolor=blue, pdfborder={0 0 0}]{hyperref}
\usepackage{enumerate}
\usepackage{color}
\usepackage[linesnumbered,boxed,ruled,commentsnumbered]{algorithm2e}
\usepackage{subfigure}
\usepackage{comment}
\usepackage{cite}
\usepackage{flushend} 
\usepackage{caption}
\usepackage{stackengine}
\newtheorem{mydef}{Definition}
\newtheorem{mythm}{Theorem}
\newtheorem{myprob}{Problem}
\newtheorem{mylem}{Lemma}

\newtheorem{mypro}{Proposition}

\newtheorem{remark}{Remark}
\usepackage[margin=1in]{geometry}

\begin{document}
\title{Safe Perception-Based Control under Stochastic Sensor \\Uncertainty using Conformal Prediction}
\author{Shuo Yang, George J. Pappas, Rahul Mangharam, and Lars Lindemann
\thanks{S. Yang, G. J. Pappas, and R. Mangharam are with the Department
of Electrical and Systems Engineering, University of Pennsylvania, Philadelphia, PA 19104, USA. L. Lindemann is with the Department of Computer Science, University of Southern California, Los Angeles, CA 90089, USA. Email: {\tt\small\{yangs1, pappasg, rahulm\}@seas.upenn.edu}, {\tt\small llindema@usc.edu}}
}
\date{}
\maketitle

\begin{abstract}
We consider perception-based control using state estimates that are obtained from high-dimensional sensor measurements via learning-enabled perception maps. However, these  perception maps are not perfect and result in state estimation errors that can lead to unsafe system behavior. Stochastic sensor noise can make matters worse and result in estimation errors that follow unknown distributions. We propose a perception-based control framework that i) quantifies estimation uncertainty of  perception maps, and ii) integrates these uncertainty representations into the control design. To do so, we use conformal prediction to compute valid state estimation regions, which are sets that contain the unknown state with high probability. We then devise a sampled-data controller for continuous-time  systems based on the notion of measurement robust control barrier functions. Our controller uses idea from self-triggered control and enables us to avoid using stochastic calculus. Our framework is agnostic to the choice of the perception map, independent of the noise distribution, and to the best of our knowledge the first to provide probabilistic safety guarantees in such a setting. We demonstrate the effectiveness of our proposed perception-based controller for a LiDAR-enabled F1/10th car.
\end{abstract}

\section{Introduction}

Perception-based control has received much attention lately~\cite{tang2018aggressive, lin2018autonomous, zhou2022safe,kantaros2022perception}. System states are usually not directly observable  and  can only be estimated from complex and noisy sensors, e.g., cameras or LiDAR. Learning-enabled perception maps can be utilized to estimate the system's state from such high-dimensional measurements. However, these estimates are usually imperfect and may lead to estimation errors, which are detrimental to the system safety.

The above observation calls for perception-based control with safety guarantees as it is crucial for many autonomous and robotic systems like self-driving cars. Recent work has been devoted to addressing these safety concerns while applying perception-based control using perception maps, see, e.g., ~\cite{dean2020guaranteeing, sun2022learning, cosner2022self, zhou2022safe, cosner2021measurement}. These work, however, either assume simple or no sensor noise models, consider specific perception maps, or lack end-to-end safety guarantees.
In realistic settings, stochastic sensor noise may be unknown and follow skewed and complex distributions that do not resemble a Gaussian distribution, as is often assumed. Additionally, perception maps can be complex, e.g., deep neural networks, making it difficult to quantify estimation uncertainty.

In this paper, we study perception-based control under stochastic sensor noise that follows arbitrary and unknown distributions. To provide rigorous safety guarantees, we have to account for estimation uncertainty caused by i) imperfect learning-enabled perception maps, and ii) noisy sensor measurements.
As shown in Figure~\ref{fig:framework}, to perform safety-critical control, we first leverage conformal prediction~\cite{vovk2005algorithmic}, a statistical tool for uncertainty quantification, to obtain state estimation regions that are valid with high probability. We then integrate these uncertain state estimation regions into the control design inspired by  the notion of measurement robust control barrier functions from~\cite{dean2020guaranteeing}.
Specifically, we design a sampled-data controller using idea from self-triggered control to ensure safety for continuous-time systems while avoiding the use of stochastic calculus.

To summarize, we make the following contributions:
\begin{itemize}
    \item We use conformal prediction to quantify  state estimation uncertainty of complex learning-enabled perception maps under arbitrary sensor and noise models;
    \item We use these uncertainty quantifications to design a sampled-data controller for continuous-time systems. We provide probabilistic safety guarantees which, to our knowledge, is the first work to do so in such a setting;
    \item We demonstrate the effectiveness of our framework in the LiDAR-enabled F1/10th vehicle simulations.
\end{itemize}

\begin{figure*}\label{fig:framework}
    \centering
    \includegraphics[width=16cm]{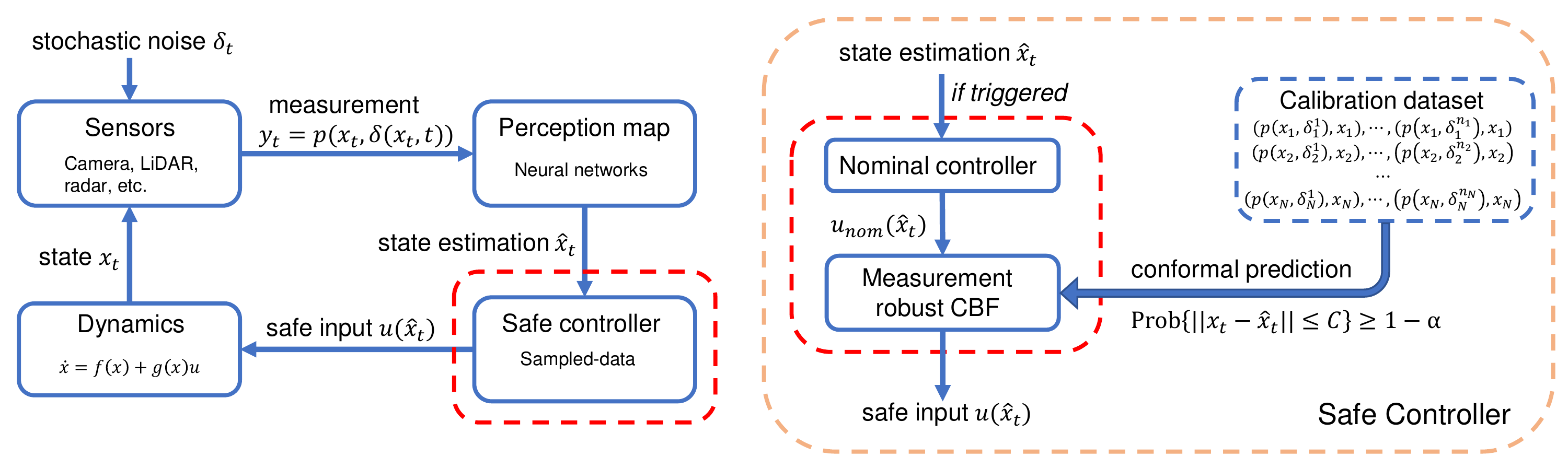}
    \caption{Overview of the system and robust safe controller. The stochastic sensor noise and imperfect perception module result in state estimation error. Conformal prediction is used to obtain the estimation error upper bound, which is then integrated into the sampled-data safe controller.}
    \label{fig:framework}
\end{figure*}

\section{Related Work}

\textbf{Perception-based control:} Control from high-dimensional sensor measurements such as cameras or LiDAR has lately gained attention. While empirical success has been reported, e.g., \cite{tang2018aggressive,lin2018autonomous,abu2021feedback, sun2023benchmark}, there is a need for designing safe-by-construction perception-enabled systems. Resilience of perception-enabled systems to sensor attacks has been studied in \cite{khazraei2022attacks,khazraei2022resiliency}, while control algorithms that provably generalize to novel environments are proposed in \cite{veer2021probably,majumdar2021pac}. In another direction, the authors in \cite{ostertag2022trajectory} plan trajectories that actively reduce estimation uncertainty. 

\textbf{Control barrier functions under estimation uncertainty:}  
Control barrier functions (CBFs) have been widely used for autonomous systems since system safety can be guaranteed~\cite{ames2016control, yang2022differentiable, xiao2023barriernet, lindemann2018control, wang2023multi}. For example, an effective data-driven approach for synthesizing safety controllers for unknown dynamic systems using CBFs is proposed in~\cite{chen2023data}.
Perception maps are first presented in combination with measurement-robust control barrier functions in \cite{dean2020guaranteeing, cosner2021measurement} when true system states are not available but only imprecise measurements. In these works, the perception error is quantified for the specific choice of the Nadarya-Watson regressor. Our approach is agnostic to the perception map and, importantly, allows to consider arbitrary stochastic sensor noise which poses challenges for continuous-time control. Measurement robust control barrier functions are learned in different variations in \cite{lindemann2021learning,dawson2022learning,sun2022learning}. Perception maps are further used to design sampled-data controllers \cite{cothren2022online,cothren2022perception,agrawal2022safe} without explicit uncertainty quantification of the sensor and perception maps.

The works in  \cite{clark2019control,clark2021control} consider state observers, e.g., extended Kalman filters, for barrier function-based control of stochastic systems. On the technical level, our approach is different as we avoid dealing with It$\hat{o}$ calculus using sampled-data control. Similarly, bounded state observers were considered in \cite{wang2022observer,zhang2022control}. However,  state observer-based approaches are generally difficult to use in perception-systems as models of high-dimensional sensors are difficult to obtain.  The authors in \cite{chou2022safe} address this challenge by combining perception maps and state observers. However, the authors assume a bound on the sensor noise and do not explicitly consider the effect of stochastic noise distributions. 

Uncertainty quantification of perception maps is vital. In similar spirit to our paper, \cite{cosner2022self,romer2022uncertainty} use (self-)supervised learning for uncertainty quantification of vision-based systems. While success is empirically demonstrated, no formal guarantees are provided as we pursue in this paper.

\textbf{Conformal prediction for control:}  Conformal prediction is a statistical method that provides probabilistic guarantees on prediction errors of machine learning models. It has been applied in computer vision~\cite{angelopoulos2020uncertainty, angelopoulos2022image}, protein design~\cite{fannjiang2022conformal}, and system verification~\cite{bortolussi2019conformal, cairoli2021neural}.
Recently, there are works that use conformal prediction for safe planning in dynamic environments, e.g.,~\cite{lindemann2022safe, dixit2022adaptive}. However, conformal prediction is only used for quantifying the prediction and not perception uncertainty, as we do in this work. To our knowledge, our work is the first to integrate uncertainty quantification from conformal prediction into perception-based control.

\section{Preliminaries and Problem Formulation}
We denote by $\mathbb{R}$, $\mathbb{N}$, and $\mathbb{R}^n$ the set of real numbers, natural numbers, and real vectors, respectively. Let $\beta$: $\mathbb{R}\rightarrow \mathbb{R}$ denote an extended class $\mathcal{K}_{\infty}$ function, i.e., a strictly increasing function with $\beta(0)=0$.
For a vector $v\in \mathbb{R}^n$, let $\|v\|$ denote its Euclidean norm.

\subsection{System Model}
We consider nonlinear control-affine systems of the form
\begin{align}\label{eq:system}
    \dot{x}(t) &= f(x(t)) + g(x(t))u(t) =: F(x(t), u(t))
\end{align}
where $x(t)\in \mathbb{R}^n$ and $u(t)\in\mathcal{U}$ are the state and the control input at time $t$, respectively, with $\mathcal{U}\subseteq \mathbb{R}^m$ denoting the set of permissible control inputs. The functions $f:\mathbb{R}^n\to\mathbb{R}^n$ and $g:\mathbb{R}^n\to\mathbb{R}^{n\times m}$ describe the internal and input dynamics, respectively, and are assumed to be locally Lipschitz continuous. 
We  assume that the dynamics in~(\ref{eq:system}) are bounded, i.e., that there exists an upper bound $\Bar{F}$ such that $\|F(x,u)\|\le \Bar{F}$ for every $(x,u)\in\mathbb{R}^n\times\mathcal{U}$.
For an initial condition $x(0)\in \mathbb{R}^n$ and a piecewise continuous control law $u:\mathbb{R}_{\ge 0}\to \mathbb{R}^m$, we denote the unique solution to the system in \eqref{eq:system} as $x:\mathcal{I}\to \mathbb{R}^n$ where $\mathcal{I}\subseteq \mathbb{R}_{\ge 0}$ is the maximum time interval on which the solution $x$ is defined.

In this paper, we assume that we do not have knowledge of  $x(t)$ during testing time, but that we observe potentially high-dimensional measurements $y(t)\in\mathbb{R}^l$ via an unknown locally Lipschitz continuous senor map $p:\mathbb{R}^n\!\times\mathbb{R}^d\!\to\!\mathbb{R}^l$ as
\begin{align}\label{eq:system_}
    y(t) &= p\big(x(t),\delta(x(t),t)\big),
\end{align}
where $\delta(x(t),t)$ is a disturbance modeled as a state-dependent random variable that is drawn from an unknown distribution $\mathcal{D}_x$ over $\mathbb{R}^d$, i.e., $\delta(x,t)\sim \mathcal{D}_{x}$.\footnote{To increase readability, we omit time indices when there is no risk of ambiguity, i.e., in this case we mean $\delta(x(t),t)\sim \mathcal{D}_{x(t)}$.} A special case that  equation \eqref{eq:system_} covers is those imperfect and noisy sensors that can be modeled as $y(t)=x(t)+\delta(t)$,  e.g., as considered in \cite{kantaros2022perception, bennetts2013towards}. The function $p(x,\delta(x,t))$ can also encode a simulated image plus noise emulating a real camera. In general, the function $p$ can model high-dimensional sensors such as camera images or LiDAR point clouds. A common assumption in recent work that we adopt implicitly in this paper is that there exists a hypothetical inverse sensor map $q:\mathbb{R}^l\to\mathbb{R}^n$ that can recover the state $x$ as $q(p(x,0))=x$ when there is no disturbance \cite{dean2020guaranteeing,dean2020robust}. This inverse sensor map $q$ is, however, rarely known and hard to model. One can instead learn perception map $\hat{q}:\mathbb{R}^l\to\mathbb{R}^n$ that approximately recovers the state $x$ such that $\|q(y,0)-\hat{q}(y)\|$ is small and bounded, which can then be used for control design \cite{dean2020guaranteeing,dean2020robust,chou2022safe}.  Note that learning an approximation of $p$ is much harder than learning the approximation $\hat{q}$ of $q$ when $l\gg n$.
\begin{remark}
The assumption on the existence of an inverse map $q$ is commonly made, as in \cite{dean2020guaranteeing,dean2020robust,chou2022safe}, and realistic when the state $x$ consists of positions and orientations that can, for instance, be recovered from a single camera image. If the state $x$ additionally consists of other quantities such as velocities, one can instead assume that $q$ partially recovers the state as $q(p(x,0))=Cx$ for a selector matrix $C$ while using a contracting  Kalman filter to estimate the remaining states when the system is detectable \cite{chou2022safe}. For the sake of simplicity, we leave this consideration for future work.  
\end{remark}

Based on this motivation, we assume that we have obtained such a perception map $\hat{q}:\mathbb{R}^l\to\mathbb{R}^n$ that estimates our state $x(t)$ at time $t$ from measurements $y(t)$, and is denoted as
\begin{align*}
    \hat{x}(t):=\hat{q}(y(t)).
\end{align*}
Note that $\hat{q}$ could be any state estimator, such as a convolutional neural network.
In our case study, we used a multi-layer perceptron (MLP) as the estimator.

\subsection{Safe Perception-Based Control Problem}

We are interested in designing control inputs $u$ from measurements $y$ that guarantee safety with respect to a continuously differentiable constraint function $h:\mathbb{R}^n\to\mathbb{R}$, i.e., so that $h(x(t))\ge 0$ for all $t> 0$ if initially $h(x(0))\ge 0$. Safety here can be framed as the controlled forward invariance of the system \eqref{eq:system} with respect to the safe set $\mathcal{C}:=\{x\in \mathbb{R}^n|h(x)\ge 0\}$ which is the superlevel set
of the function $h$. The difficulty in this paper is that we are not able to measure the state $x(t)$ directly during runtime, and that we have only sensor measurements $y(t)$ from the unknown and noisy sensor map $p$ available.
\begin{myprob}\label{problem}
Consider the system in \eqref{eq:system} with initial state $x(0)\in\mathbb{R}^n$ and sensor model in \eqref{eq:system_}. Let $h:\mathbb{R}^n\to\mathbb{R}$ be a continuously differentiable constraint function, $\mathcal{T}\subset\mathbb{R}_{\ge 0}$ be a time interval, and $\alpha$ be a failure probability. Design a control input $u$ from sensor measurements $y$ such that $\text{Prob}(x(t)\in\mathcal{C}, \forall t\in \mathcal{T})\ge 1-\alpha$.
\end{myprob}

\subsection{Uncertainty Quantification via Conformal Prediction}

In our solution to Problem \ref{problem}, we use conformal prediction which is a statistical tool introduced  in \cite{vovk2005algorithmic,shafer2008tutorial} to obtain valid uncertainty regions for complex prediction models without making assumptions on the underlying distribution or the prediction model \cite{angelopoulos2021gentle,lei2018distribution}.  Let $Z,Z^{(1)},\hdots,Z^{(k)}$ be $k+1$ independent and identically distributed real-valued random variables, known as the nonconformity scores. Our goal is to obtain an uncertainty region for $Z$ defined via a function $\bar{Z}:\mathbb{R}^k\to\mathbb{R}$  so that $Z$ is bounded by $\bar{Z}(Z^{(1)},\hdots,Z^{(k)})$ with high probability.  Formally, given a failure probability $\alpha\in (0,1)$, we want to construct an  uncertainty region $\bar{Z}$ such that $\text{Prob}(Z\le \bar{Z})\ge 1-\alpha$ where we  omitted the dependence of $\bar{Z}$ on $Z^{(1)},\hdots,Z^{(k)}$ for convenience. 

By a surprisingly simple quantile argument, see \cite[Lemma 1]{tibshirani2019conformal}, the uncertainty region $\bar{Z}$ is obtained as the $(1-\alpha)$th quantile of the empirical distribution over the values of $Z^{(1)},\hdots,Z^{(k)}$ and $\infty$. We recall this result next.
\begin{mylem}[Lemma 1 in \cite{tibshirani2019conformal}]\label{lem:1}
Let $Z,Z^{(1)},\hdots,Z^{(k)}$ be $k+1$ independent and identically distributed real-valued random variables. Without loss of generality, let $Z^{(1)},\hdots,Z^{(k)}$ be sorted in non-decreasing order and define $Z^{(k+1)}:=\infty$. For $\alpha\in(0,1)$, it holds that $\text{Prob}(Z\le \bar{Z})\ge 1-\alpha$ where
\begin{align*}
    \bar{Z}:=Z^{(r)} \text{ with } r:=\lceil (k+1)(1-\alpha)\rceil
\end{align*}
and where $\lceil\cdot\rceil$ is the ceiling function.
\end{mylem} 
 Some clarifying comments are in order. First, we remark that $\text{Prob}(Z\le \bar{Z})$ is a marginal probability over the randomness in $Z,Z^{(1)},\hdots,Z^{(k)}$ and not a conditional probability. Second, note that $\lceil (k+1)(1-\alpha)\rceil>k$ implies that $\bar{Z}=\infty$.

\section{Safe Perception-Based Control with Conformal Prediction}

Addressing Problem \ref{problem} is challenging for two reasons. First, the perception map $\hat{q}$ may not be exact, e.g., even in the disturbance-free case, it may not hold that $\hat{q}(p(x,0))=x$. Second, even if we have accurate state estimates in the disturbance-free case, i.e., when $\hat{q}(p(x,0))$ is close to $x$, this does not imply that we have the same estimation accuracy with disturbances, i.e., $\hat{q}(p(x,\delta))$ may not necessarily be close to $x$. Our setting is thus distinctively different from existing approaches and requires uncertainty quantification of the noisy error between $\hat{x}(t)$ and $x(t)$. 

\subsection{Conformal Prediction for Perception Maps}
Let us now denote the stochastic state estimation error as 
\begin{align*}
    e(x,t):=\|\hat{x}-x\|=\|\hat{q}\big(\underbrace{p(x,\delta(x,t))}_{=y}\big)-x\|.
\end{align*}
For a fixed state $x\in\mathbb{R}^n$, our first goal is  to construct a prediction region $\bar{E}_x$  so that
\begin{align}\label{eq:nonconf_e}
    \text{Prob}\big(e(x,t)\le \bar{E}_x\big)\ge 1-\alpha
\end{align}
holds uniformly over $t\in\mathbb{R}_{\ge 0}$. Note that the distribution $\mathcal{D}_x$ of $\delta$ is independent of time $t$ so that we will get uniformity automatically.  While we do not know the sensor map $p$, we assume here that we have an oracle that gives us $N\ge \lceil (N+1)(1-\alpha)\rceil$ state-measurement data pairs $(x,y^{(i)})$ called calibration dataset, where $i\in\{1,\hdots,N\}$ and $y^{(i)}=p(x,\delta^{(i)})$ with $\delta^{(i)}\sim\mathcal{D}_x$. This is a common assumption, see, e.g., \cite{chou2022safe,dean2020guaranteeing}, and such an oracle can, for instance, be a simulator that we can query data from. By defining the nonconformity score $Z^{(i)}:=\|\hat{q}(y^{(i)})-x\|$, and assuming that $Z^{(i)}$ are sorted in non-decreasing order, we can now obtain the guarantees in equation \eqref{eq:nonconf_e} by applying Lemma \ref{lem:1}. In other words, we obtain $\bar{E}_x:=Z^{(r)}$ with $r$ from Lemma \ref{lem:1} so that $\text{Prob}\big(\hat{x}\in\{\zeta\in\mathbb{R}^n|\|x-\zeta\|\le\bar{E}_x\}\big)\ge 1-\alpha$ holds. Note that this gives us information about the estimate $\hat{x}$, but not about the state $x$ which was, in fact, fixed a-priori. To revert this argument and obtain a prediction region for $x$ from $\hat{x}$, we have to ensure that equation \eqref{eq:nonconf_e} holds for a set of states instead of only a single state $x$, which will be presented next. To do so, we use a covering argument next.

 Consider now a compact subset of the workspace $\mathcal{X}\subseteq \mathbb{R}^n$ that should include the safe set $\mathcal{C}$. Let  $\epsilon>0$ be a gridding parameter and construct an $\epsilon$-net $\bar{\mathcal{X}}$ of $\mathcal{X}$, i.e., construct a finite set $\bar{\mathcal{X}}$  so that for each $x\in\mathcal{X}$ there exists an $x_j\in \bar{\mathcal{X}}$ such that $\|x-x_j\|\le \epsilon$. For this purpose, simple gridding strategies can be used as long as the set $\mathcal{X}$ has a convenient representation. Alternatively, randomized algorithms can be used that sample from $\mathcal{X}$ \cite{vershynin2018high}. We can now again apply a conformal prediction argument for each grid point $x_j\in\bar{\mathcal{X}}$ and then show the following proposition.

\begin{mypro}\label{prop:bound}
Consider the Lipschitz continuous sensor map $p$ in \eqref{eq:system_} and a perception map $\hat{q}$ with respective Lipschitz constants $\mathcal{L}_p$ and $\mathcal{L}_{\hat{q}}$.\footnote{We assume that the Lipschitz constant of the sensor map $p$ is uniform over the parameter $\delta$, i.e., that $\delta$ does not affect the value of $\mathcal{L}_{p}$.} Assume that we constructed an $\epsilon$-net $\bar{\mathcal{X}}$ of $\mathcal{X}$. For each $x_j\in\bar{\mathcal{X}}$, let $(x_j,y_j^{(i)})$ be $N\ge \lceil (N+1)(1-\alpha)\rceil$ data pairs where $y_j^{(i)}:=p(x_j,\delta^{(i)})$ with $\delta^{(i)}\sim \mathcal{D}_{x_j}$. Define $Z_j^{(i)}:=\|\hat{q}(y_j^{(i)})-x_j\|$, and assume that $Z_j^{(i)}$ are sorted in non-decreasing order, and let $\bar{E}_{x_j}:=Z^{(r)}_j$ with $r$ from Lemma \ref{lem:1}. Then, for any $x\in\mathcal{X}$, it holds that
\begin{align}\label{ineq:estimation-bound}
    \text{Prob}\Big(e(x,t)\le\sup_j\bar{E}_{x_j}+(\mathcal{L}_{p}\mathcal{L}_{\hat{q}}+1)\epsilon\Big)\ge 1-\alpha,
\end{align}
\end{mypro}
\begin{proof}
See Appendix.
\end{proof}

The above result says that the state estimation error $e(x,t)$ can essentially be bounded, with probability $1-\alpha$, by the worst case of conformal prediction region $\bar{E}_{x_j}$ within the grid $\bar{\mathcal{X}}$ and by the gridding parameter $\epsilon$. Under the assumption that our system operates in the workspace $\mathcal{X}$ and based on inequality \eqref{ineq:estimation-bound}, we can hence conclude that \begin{align*}
    \text{Prob}\Big(\!x\in\!\{\zeta\!\in\!\mathbb{R}^n|\|\zeta\!-\!\hat{x}\|\!\le\! \sup_j\bar{E}_{x_j}\!\!+\!(\mathcal{L}_{p}\mathcal{L}_{\hat{q}}\!+\!1)\epsilon\}\!\Big)\ge\! 1-\alpha.
\end{align*}
\begin{remark}
We note that the Lipschitz constants of the sensor and perception maps are used in the upper bound in \eqref{ineq:estimation-bound} (as commonly done in the literature \cite{dean2020guaranteeing,cothren2022online,chou2022safe}), which may lead to a conservative bound. One practical way to mitigate this conservatism is to decrease the gridding parameter $\epsilon$, i.e., to increase the sampling density in the workspace $\mathcal{X}$.
\end{remark}

\subsection{Sampled-Data Controller using Conformal Estimation Regions}
After bounding the state estimation error in Proposition \ref{prop:bound}, we now design a uncertainty-aware controller based on  equation \eqref{ineq:estimation-bound}. However, a technical challenge in doing so is that the measurements are stochastic. By designing a sampled-data controller, we can avoid difficulties dealing with stochastic calculus. To do so,  we first present a slightly modified version of measurement robust control barrier function (MR-CBF) introduced in~\cite{dean2020guaranteeing}.
\begin{mydef}
Let $\mathcal{C}\subseteq \mathbb{R}^n$ be the zero-superlevel set of a continuously differentiable function $h: \mathbb{R}^n\rightarrow \mathbb{R}$. The function $h$ is
a measurement robust control barrier function (MR-CBF) for the system in (\ref{eq:system}) with parameter function pair $(a, b):\mathbb{R}^l\rightarrow \mathbb{R}^2_{\ge0}$ if there exists an extended class $K_{\infty}$ function $\beta$ such that
\begin{align}\label{eq:mrcbf}
    &\sup_{u\in \mathcal{U}}[L_fh(\hat{x})+L_gh(\hat{x})u-(a(y)+b(y)\|u\|)]\ge -\beta(h(\hat{x}))
\end{align}
for all $(y, \hat{x})\in V(\mathcal{C})$, where $V(\mathcal{C}):=\{(y, \hat{x})\in \mathbb{R}^l\times \mathbb{R}^n|\exists (x,\delta)\in \mathcal{C}\times \mathcal{D}_x \text{ s.t. } \hat{x}=\hat{q}(p(x, \delta))\}$, and $L_fh(\hat{x})$ and $L_gh(\hat{x})$ denote the Lie
derivatives.
\end{mydef}
Compared to regular CBFs \cite{ames2016control}, a MR-CBF introduces a non-positive robustness term $-(a(y)+b(y)\|u\|)$ which makes the constraint in \eqref{eq:mrcbf} more strict. Now, given a MR-CBF $h(x)$, the set of MR-CBF consistent control inputs is
\begin{align}
    &K_{CBF}(y):=\{u\in \mathcal{U}|L_fh(\hat{x})+L_gh(\hat{x})u\nonumber\\
    &\quad\quad\quad\quad\quad\quad-(a(y)+b(y)\|u\|) +\beta(h(\hat{x}))\ge 0\}.
\end{align}
Note that we can not simply follow \cite[Theorem 2]{dean2020guaranteeing} to obtain a safe control law as $u(t)\in K_{CBF}(y(t))$ since $y(t)$ and consequent $u(t)$ are stochastic. We hence propose a sampled-data control law  that keeps the trajectory $x(t)$ within the set $\mathcal{C}$ with high probability.
The sampled-data control law $\hat{u}$ is piecewise continuous and defined as
\begin{align}\label{event_control}
    \hat{u}(t):=u(t_i), \;\forall t\in[t_i, t_{i+1}),
\end{align}
where $u(t_i)$ at triggering time $t_i$ is computed by solving the following quadratic optimization problem
\begin{equation} \label{eq:safety_filter}
\begin{aligned}
    u(t_{i}) &= \underset{u\in K_{CBF}(y)}{\text{argmin}}  \quad \lVert u - u_{nom}(t_i)) \rVert^2,
\end{aligned}
\end{equation}
where $u_{nom}(t_i)$ is any nominal control law that may not necessarily be safe. 
Then, we select the triggering instances $t_i$ as follows:
\begin{align}\label{self_time}
    t_0&:=0,\nonumber\\
    t_{i+1}&:=(\Delta-\sup_j\bar{E}_{x_j}-(\mathcal{L}_{p}\mathcal{L}_{\hat{q}}+1)\epsilon)/\Bar{F}+t_i,
\end{align}
where $\Delta$ is a user-defined parameter that will define the parameter pair $(a, b)$ of the MR-CBF and that has to be $\Delta>\sup_j\bar{E}_{x_j}+(\mathcal{L}_{p}\mathcal{L}_{\hat{q}}+1)\epsilon$.
Naturally, larger $\Delta$ lead to less frequent control updates, but will require more robustness and reduce the set of permissible control inputs in $K_{CBF}(y)$.
Based on the computation of triggering times in \eqref{self_time}, the following lemma holds.
\begin{mylem}\label{lem:self}
    Consider the sampled-data control law $\hat{u}(t)$ in~(\ref{event_control}) with the triggering rules~(\ref{self_time}), it holds that
    \begin{align}
        \text{Prob}\Big(\|x(t)-\hat{x}(t_i)\|\le\Delta, \forall t\in[t_i, t_{i+1})\Big)\ge 1-\alpha.
    \end{align}
\end{mylem}
\begin{proof}
    See Appendix.
\end{proof}
Intuitively, the above lemma says that $\|x(t)-\hat{x}(t_i)\|\le \Delta$ holds with high probability in between  triggering times if the sampled-data control law $\hat{u}(t)$ in~(\ref{event_control}) with the triggering rules~(\ref{self_time}) is executed.
Then, we can obtain the following probabilistic safety guarantees.
\begin{mythm}\label{thm:main_self_safety}
Consider a MR-CBF $h$ with parameter pair $(a(y), b(y))=((\mathcal{L}_{L_fh}+\mathcal{L}_{\beta\circ h})\Delta, \mathcal{L}_{L_gh}\Delta)$
where $\mathcal{L}_{L_fh}, \mathcal{L}_{\beta\circ h},$ and $\mathcal{L}_{L_gh}$ are the Lipschitz constants of the functions $L_fh, \beta\circ h$ and $L_gh$, respectively. Then, for any nominal control law $u_{nom}$, the sampled-data law $\hat{u}(t)$ in~(\ref{event_control}) with the triggering rule in~(\ref{self_time}) will render the set $\mathcal{C}$ forward invariant with a probability of at least  $1-\alpha$. In other words, we have that
\begin{align}\label{safe_guarantee_prob_event}
    \text{Prob}\Big(x(t)\in\mathcal{C}, \forall t\in[t_i, t_{i+1})\Big)\ge 1-\alpha.
\end{align}
\end{mythm}
\begin{proof}
    See Appendix.
\end{proof}
The above theorem solves  Problem~\ref{problem} for the time interval $\mathcal{T}=[t_i, t_{i+1})$. 
If we want to consider a larger time interval $\mathcal{T}=[0, T)$ under the sampled-data control law, we have the following guarantees.
\begin{mypro}\label{prop:main_self_full_guarantee}
Under the same condition as in Theorem~\ref{thm:main_self_safety}, for a time interval $\mathcal{T}=[0, T)$, we have that:
    \begin{align}\label{safe_guarantee_prob_self_power}
    \text{Prob}\Big(x(t)\in\mathcal{C}, \forall t\in[0, T)\Big)\ge (1-\alpha)^m,
\end{align}
where $m\in\mathbb{N}_{> 0}$ such that $t_{m-1}\le T<t_{m}$.
\end{mypro}
\begin{proof}
    See Appendix.
\end{proof}
Note that if we want to achieve any probability guarantee $p\in (0,1)$, we can just let $(1-\alpha)^m=p$ and obtain $\alpha=1-p^{1/m}$.

\section{Simulation Results}

\begin{figure}
    \centering
    \includegraphics[width=5.0cm]{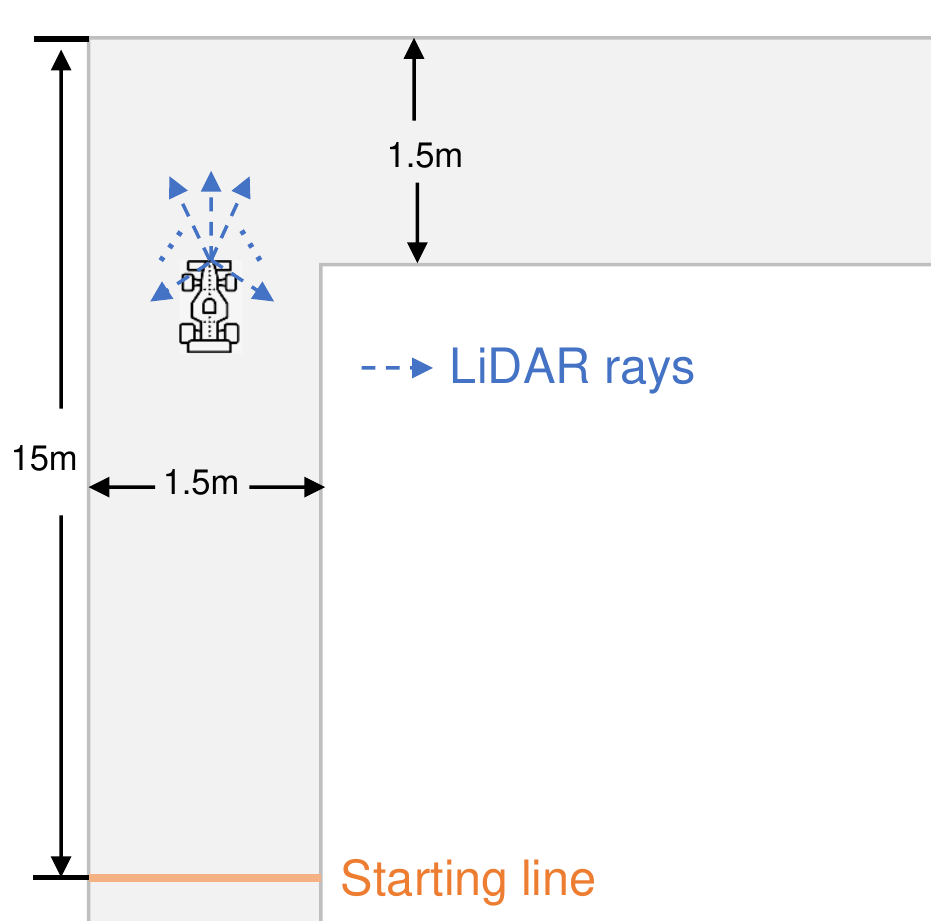}
    \caption{The F1/10 vehicle is equipped with a 2D LiDAR sensor that outputs an array of 64 laser scans. The vehicle starts at a random position on the starting line.}
    \label{fig:scenario}
\end{figure}

To demonstrate our proposed safe perception-based control law, we consider navigating an F1/10th autonomous vehicle in a structured environment~\cite{ivanov2020case}, which is shown in Figure~\ref{fig:scenario}.
The vehicle system has the state $x=[p_x, p_y, \theta]$, where $[p_x, p_y]$ denotes its position and $\theta$ denotes its orientation. We have $[\dot{p}_x, \dot{p}_y]=[u_x, u_y]$, where $u_x$ and $u_y$ are control inputs denoting velocities, and $\theta=\arctan(u_y/u_x)$.
The control input constraint is $(u_x, u_y)\in [-1, 1]\times[-1, 1]$.
Thus, the assumption that there exists an upper bound $\Bar{F}$ for dynamics holds for this system.

\textbf{Observation model}:
The vehicle is equipped with a 2D LiDAR scanner from which it obtains LiDAR measurements as its observations. Specifically, the measurement include 64 LiDAR rays uniformly ranging from $-\frac{3\pi}{4}$ to $\frac{3\pi}{4}$ relative to the vehicle's heading direction.
To model the uncertainty of measurements, unknown noise conforming to exponential distribution is added to each ray:
\begin{align*}
    y^k_n = y^k+\delta, \quad \delta\sim exp(\lambda),
\end{align*}
where $y^k$ is the ground truth for ray $k$, $y^k_n$ is the corrupted observed ray $k$, and $\lambda$ is the parameter of exponential distribution, where the noise $\delta$ is drawn from.
In our experiments, we let $\lambda:=2/3$.

\begin{figure}
    \centering
    \includegraphics[width=6.3cm]{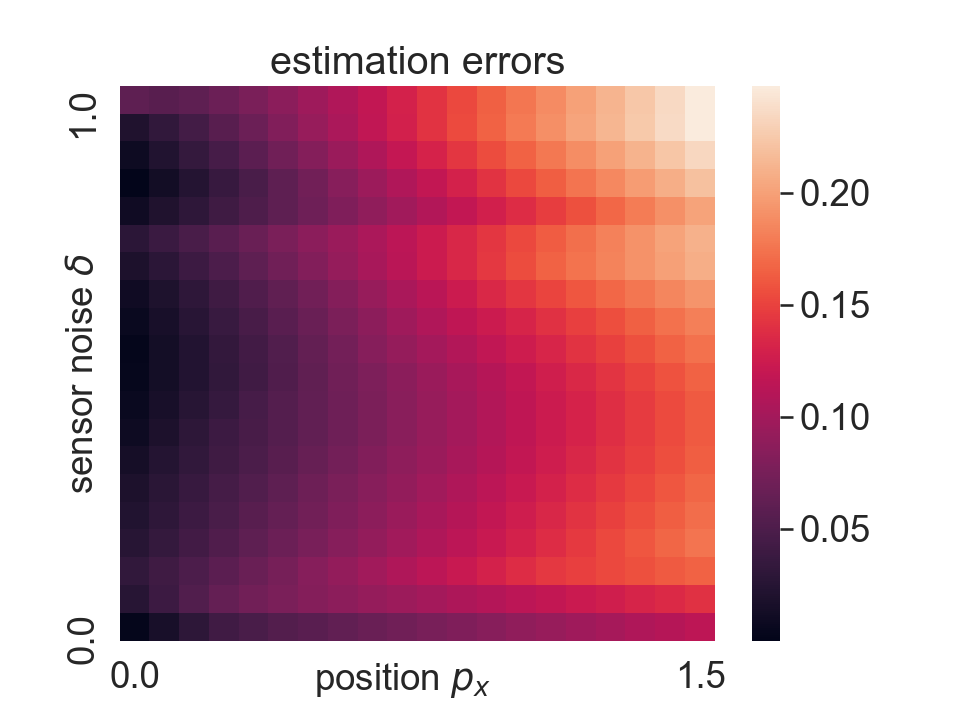}
    \caption{The empirical model errors $e(x)$ w.r.t. $p_x$ and $\delta$ measured on a validation set. $p_y$ and $\theta$ are fixed.}
    \label{fig:error}
\end{figure}

\textbf{Perception map}:
We trained a feedforward neural network to estimate the state of the vehicle.
The input is the 64-dimensional LiDAR measurement and output is the vehicle's state.
The training dataset $D_{train}$ contains $4\times 10^5$ data points, and the calibration dataset $D_{cal}$ for conformal prediction contains $1.25\times 10^4$ data points.
For illustration, under a fixed heading $\theta$ and longitudinal position $p_y$, the errors $e(x)$ of the learned perception map with respect to sensor noise $\delta$ and horizontal position $p_x$ is shown in Figure~\ref{fig:error}.

\textbf{Barrier functions:}
To prevent collision with the walls, when the vehicle is traversing the long hallway, the CBF is chosen as $h(x)=\min\{h_1(x), h_2(x)\}$, where $h_1(x)=p_x$ and $h_2(x)=1.5 - p_x$. 
Then we have the safe set $\mathcal{C}=\{x\in\mathbb{R}^3|h(x)\ge 0\}$.
CBFs can be similarly defined when the vehicle is operating in the corner.
To demonstrate the effectiveness of our method, we compare the following two cases in simulations:
\begin{enumerate}
    \item \emph{Measurement robust CBF:} as shown in Theorem~\ref{thm:main_self_safety}, we choose the parameters pair $(a(y), b(y))=((\mathcal{L}_{L_fh}+\mathcal{L}_{\beta\circ h})\Delta, \mathcal{L}_{L_gh}\Delta)$ to ensure robust safety.
    \item \emph{Vanilla CBF:} we choose the parameters pair $(a(y), b(y)) = (0, 0)$, which essentially reduces to the vanilla non-robust CBF~\cite{ames2016control}.
    However, the perceived state is from perceptual estimation rather than real state, so this CBF cannot provide any safety guarantee.
\end{enumerate}
Note that we obtain necessary Lipschitz constants using sampling-based estimation method in simulations.

\begin{figure}
    \centering
    \includegraphics[width=6cm]{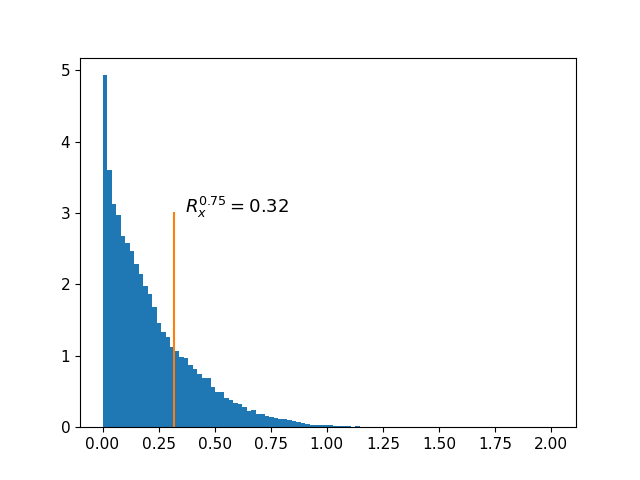}
    \caption{Nonconformity scores $R_x$ histogram during runtime. We select the coverage rate as $75\%$. }
    \label{fig:score}
\end{figure}

\begin{figure}
    \centering
    \subfigure[Measurement robust CBF]{\label{fig:robustTraj}
        \includegraphics[width=0.26\textwidth]{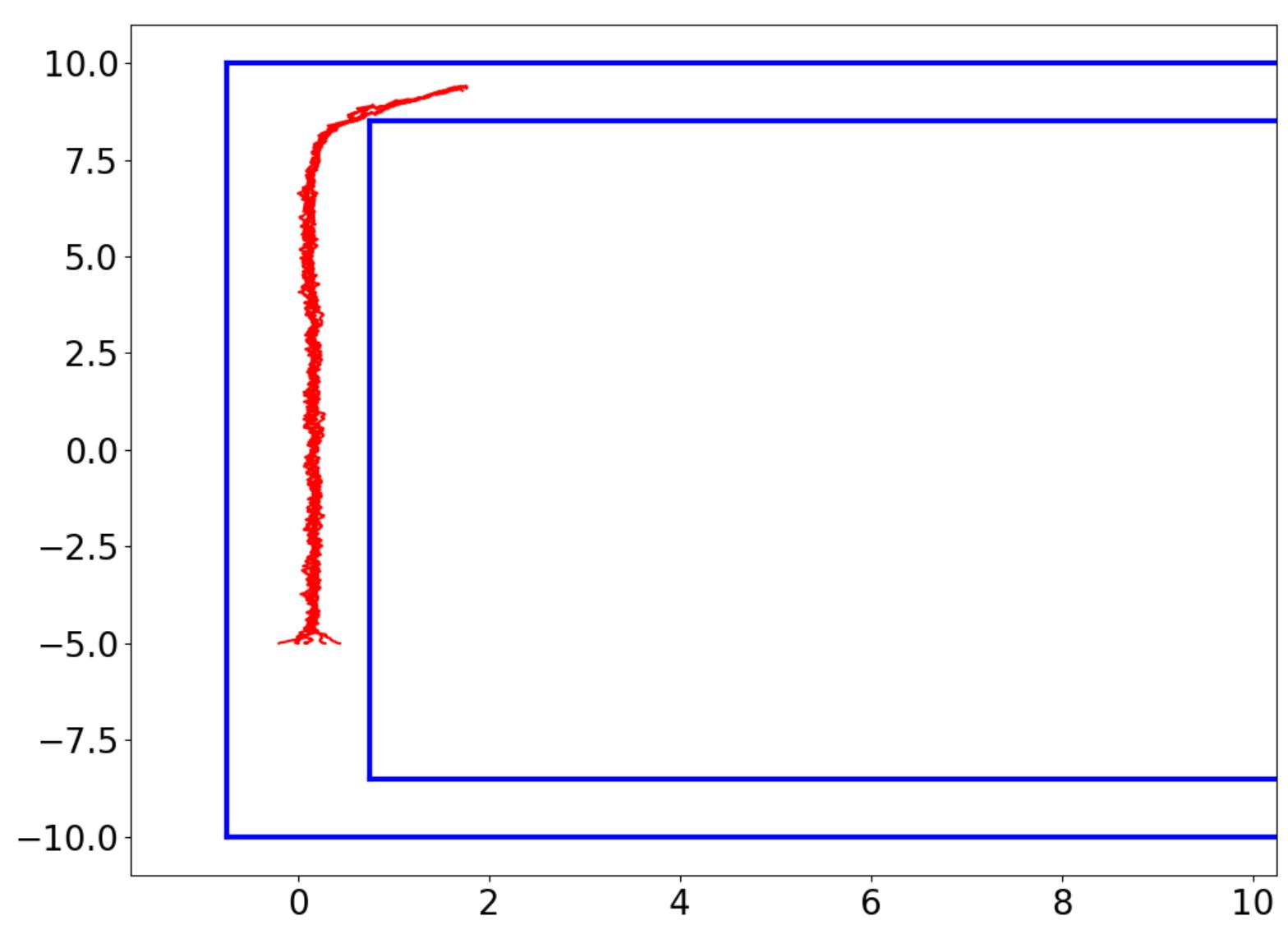}}
    \subfigure[Vanilla CBF]{\label{fig:nonrobustTraj}
        \includegraphics[width=0.26\textwidth]{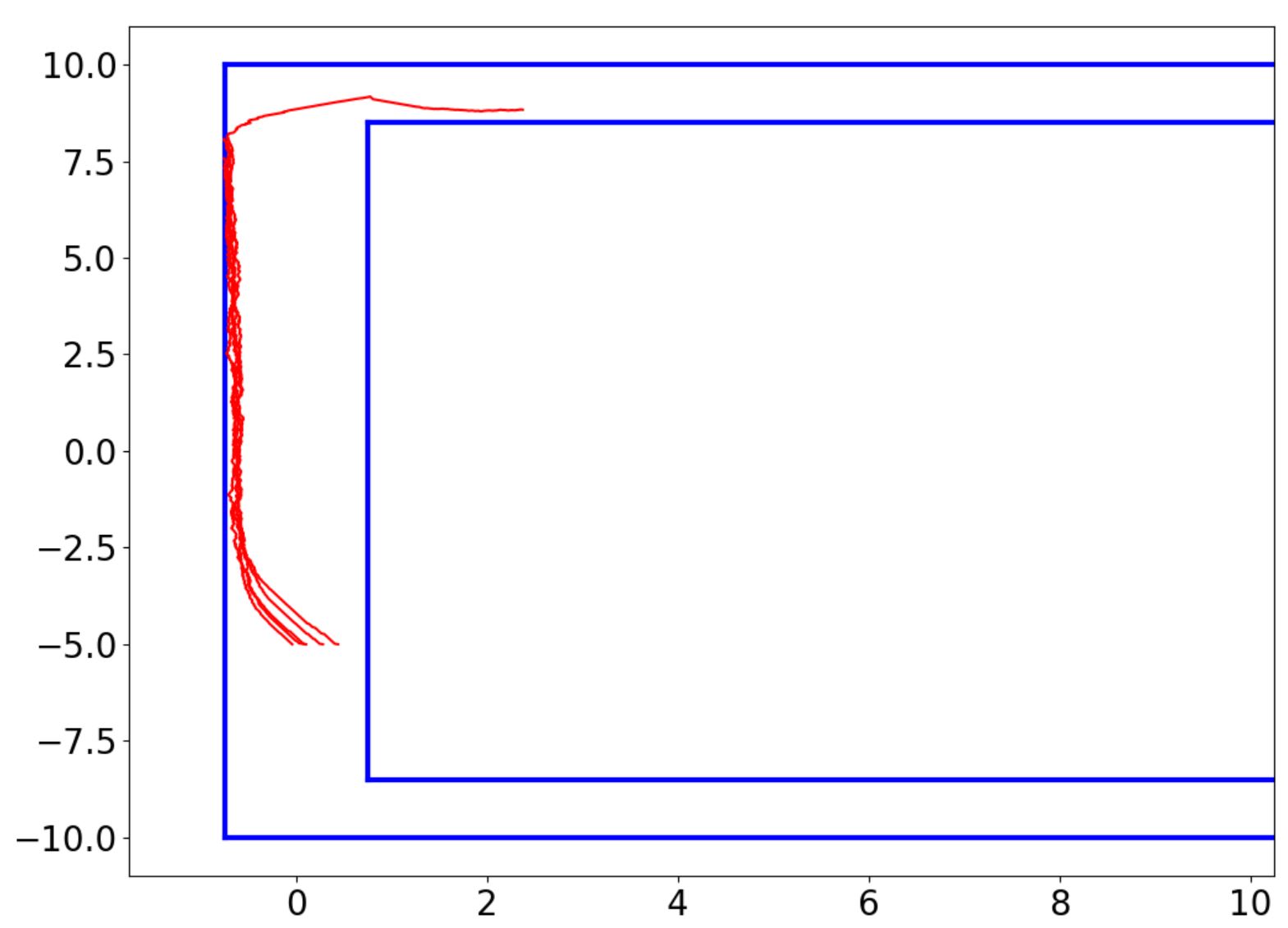}}
    \caption{Traces for the sampled-data measurement robust CBF and vanilla CBF (5 traces are presented). All traces are tested with horizon $T=30s$. We run 100 traces totally, and the safety rates are 93\% and 16\%, respectively.}. \label{fig:traj}
\end{figure}

\textbf{Uncertainty and results}:
The vehicle is expected to track along the hallway and make a successful turn in the corner as shown in Figure~\ref{fig:scenario}. 
The nominal controller is a PID controller.
We set the coverage error $\alpha=0.25$, so we desire 
$P(\|x_t-\hat{x}_t\|\le \epsilon')\ge 1-\alpha=75\%$.
Based on calibration of conformal prediction and Proposition~\ref{prop:bound}, we calculate that $\epsilon'=0.34$, and we choose $\Delta=0.35>\epsilon'$.
The nonconformity score histogram is presented in Figure~\ref{fig:score}, in which the $75\%$ quantile value is $R_x^{0.75}=0.32<\epsilon'$, so our Proposition~\ref{prop:bound} holds in practice.
As presented in Figure~\ref{fig:traj}, the safety rate of sampled-data measurement robust CBF is $93\%$, which is significantly higher than vanilla non-robust CBF case ($16\%$).

\section{Conclusion}
In this paper, we consider the safe perception-based control problem under stochastic sensor noise.
We use conformal prediction to quantify the state estimation uncertainty, and then integrate this uncertainty into the design of sampled-data safe controller.
We obtain probabilistic safety guarantees for continuous-time systems.
Note that, in this work, the perception map only depends on current observation, which might limit its accuracy in some cases. We plan to incorporate history observations into perception maps in the future.
Also, we are interested in providing a more sample-efficient scheme while constructing calibration dataset.


\bibliographystyle{IEEEtran}
\bibliography{reference}
\section{Appendix}

\emph{Proof of Proposition~\ref{prop:bound}}: First, note that
\begin{align*}
    &\|p(x, \delta) - p(x', \delta)\|\le \mathcal{L}_p\|x-x'\|, \\
    &\|\hat{q}(p(x, \delta) - p(x', \delta))\|\le \mathcal{L}_{\hat{q}}\cdot\|(p(x, \delta) - p(x', \delta)\|.
\end{align*}
due to Lipschitz continuity of $p$ and $\hat{q}$. Since, for any $x\in\mathcal{X}$, there exists a $x_i\in \bar{\mathcal{X}}$ such that $\|x-x_i\|\le \epsilon$, we know that
\begin{align*}
    \|\hat{x}-\hat{x}_i\|&=\|\hat{q}(p(x, \delta))-\hat{q}(p(x_i, \delta))\|\le \mathcal{L}_{\hat{q}}\|p(x, \delta), p(x_i, \delta)\|\\
    &\le \mathcal{L}_{\hat{q}}\mathcal{L}_{p}\|x-x_i\|\le \mathcal{L}_{\hat{q}}\mathcal{L}_{p}\epsilon.
\end{align*}
Thus, we can bound the state estimation error $e(x, t)$ with probabability at least $1-\alpha$ as
\begin{align*}
    e(x, t)&=\|\hat{x}-x\|=\|\hat{x}-x+\hat{x}_i-\hat{x}_i+x_i-x_i\|\\
    &\le \|\hat{x}-\hat{x}_i\|+\|x_i-x\|+\|\hat{x}_i-x_i\|\\
    &\le \mathcal{L}_{\hat{q}}\mathcal{L}_{p}\epsilon+\epsilon+\bar{E}_{x_i}\le (\mathcal{L}_{\hat{q}}\mathcal{L}_{p}+1)\epsilon+\sup_j\bar{E}_{x_j}.
\end{align*}
Particularly, note that the last inequality holds since $\text{Prob}(\|\hat{x}_i-x_i\|\le \bar{E}_{x_i})\ge 1-\alpha$ for each $x_i\in\Bar{\mathcal{X}}$. Finally, we have that $\text{Prob}\big(e(x,t)\le\sup_j\bar{E}_{x_j}+(\mathcal{L}_{p}\mathcal{L}_{\hat{q}}+1)\epsilon\big)\ge 1-\alpha$.

\emph{Proof of Lemma~\ref{lem:self}:}
First, recall the system dynamics:
\[F(x(t), u(t)):=f(x(t))+g(x(t))u(t)=\dot{x}(t).\]
By integrating the above ODE, we have that
\begin{align}
    x(t)=x(t_i)+\int_{t_i}^{t} F(x(s), u(s)) \,ds.\nonumber
\end{align}
Then, for any $t\in[t_i, t_{i+1})$, it holds w.p. $1-\alpha$ that
\begin{align}
    \|x(t)-\hat{x}(t_i)\|&= \|\int_{t_i}^{t} F(x(s), u(s)) \,ds\|+\|x(t_i)-\hat{x}(t_i)\|\nonumber\\
    &\hspace{-1cm}\le (t-t_i)\Bar{F}+\sup_j\bar{E}_{x_j}+(\mathcal{L}_{p}\mathcal{L}_{\hat{q}}+1)\epsilon\le\Delta,
\end{align}
where we used that Prob$(\|x(t_i)-\hat{x}(t_i)\|\le \sup_j\bar{E}_{x_j}+(\mathcal{L}_{p}\mathcal{L}_{\hat{q}}+1)\epsilon)\ge 1-\alpha$ according to Proposition~\ref{prop:bound}.
Thus, we  have that $\text{Prob}\Big(\|x(t)-\hat{x}(t_i)\|\le\Delta, \forall t\in[t_i, t_{i+1})\Big)\ge 1-\alpha.$

\emph{Proof of Theorem~\ref{thm:main_self_safety}}:
Let us first define
\begin{align}
    &c(x(t), u(t)):=L_fh(x(t))+L_gh(x(t))u(t)+\beta(h(x(t))),\nonumber\\
    &c(\hat{x}(t), u(t)):=L_fh(\hat{x}(t))+L_gh(\hat{x}(t))u(t)+\beta(h(\hat{x}(t))).
\end{align}
For any $t\in[t_i, t_{i+1})$, we can now upper bound
the absolute difference between $c(\hat{x}(t_i), u(t_i))$ and $c(x(t), u(t))$ as
\begin{align}
    &|c(\hat{x}(t_i), u(t_i))-c(x(t), u(t))|\nonumber\\
    =&|(L_fh(\hat{x}(t_i))+L_gh(\hat{x}(t_i))u(t_i)+\beta(h(\hat{x}(t_i))))\nonumber\\
    &\quad-(L_fh(x(t))+L_gh(x(t))u(t)+\beta(h(x(t))))|\nonumber\\
    =&|(L_fh(\hat{x}(t_i))-L_fh(x(t)))+(L_gh(\hat{x}(t_i))u(t_i)\nonumber\\
    &\quad-L_gh(x(t))u(t))+(\beta(h(\hat{x}(t_i)))-\beta(h(x(t))))|\nonumber\\
    \le &(\mathcal{L}_{L_fh}+\mathcal{L}_{L_gh}\|u(t_i)\|+\mathcal{L}_{\beta\circ h})\cdot\|x(t)-\hat{x}(t_i)\|\nonumber\\
    \le &(\mathcal{L}_{L_fh}+\mathcal{L}_{L_gh}\|u(t_i)\|+\mathcal{L}_{\beta\circ h})\cdot\Delta \quad\text{(w.p. $1-\alpha$)}
\end{align}
Since $c(\hat{x}(t_i), u(t_i))\ge (\mathcal{L}_{L_fh}+\mathcal{L}_{L_gh}\|u(t_i)\|+\mathcal{L}_{\beta\circ h})\cdot\Delta$,
we can quickly obtain $c(x(t), u(t))\ge c(\hat{x}(t_i), u(t_i))-(\mathcal{L}_{L_fh}+\mathcal{L}_{L_gh}\|u(t_i)\|+\mathcal{L}_{\beta\circ h})\cdot\Delta\ge 0$ using the absolute difference bound we derived above, which implies $h(x(t))\ge 0$, $\forall t\in[t_i, t_{i+1})$.
Also, $\|x(t)-\hat{x}(t_i)\|\le \Delta$ holds with the probability $1-\alpha$, so we finally obtain that $\text{Prob}\Big(h(x(t))\ge 0, \forall t\in[t_i, t_{i+1})\Big)\ge 1-\alpha$. 
This ends the proof.

\emph{Proof of Proposition~\ref{prop:main_self_full_guarantee}}:
We first prove that 
\begin{align}
    &P\{x(t)\in\mathcal{C}, \forall t\in[0, t_m)\}\nonumber\\
    \ge& (1-\alpha)\cdot P\{x(t)\in\mathcal{C}, \forall t\in[t_{0}, t_{m-1})\},
\end{align}
which can be obtained by the following derivations:
\begin{align}
    &P\{x(t)\in\mathcal{C}, \forall t\in[0, t_m)\}\nonumber\\
    =&P\{x(t)\in\mathcal{C}, \forall t\in[0, t_1)\cup[t_1, t_2)\cup\cdots\cup[t_{m-1}, t_m)\}\nonumber\\
    =&P\{x(t)\in\mathcal{C}, \forall t\in[t_{m-1}, t_m)|x(t)\in\mathcal{C}, \forall t\in[t_{0}, t_{m-1})\}\nonumber\\
    &\quad\cdot P\{x(t)\in\mathcal{C}, \forall t\in[t_{0}, t_{m-1})\}\nonumber\\
    =&P\{x(t)\in\mathcal{C}, \forall t\in[t_{m-1}, t_m)|x(t)\in\mathcal{C}, \forall t\in[t_{0}, t_{m-1}]\}\nonumber\\
    &\quad\cdot P\{x(t)\in\mathcal{C}, \forall t\in[t_{0}, t_{m-1})\}\nonumber\\
    =&P\{x(t)\in\mathcal{C}, \forall t\in[t_{m-1}, t_m)|x(t_{m-1})\in\mathcal{C}\}\nonumber\\
    &\quad\cdot P\{x(t)\in\mathcal{C}, \forall t\in[t_{0}, t_{m-1})\}\nonumber\\
    \ge& (1-\alpha)\cdot P\{x(t)\in\mathcal{C}, \forall t\in[t_{0}, t_{m-1})\}
\end{align}
Note that the third equality holds due to the continuity of $h(x)$, i.e., if $\lim_{t\rightarrow t_{m-1}}h(x(t))\ge 0$, then we have that $h(x(t_{m-1}))\ge 0$,
which implies that the event $x(t)\in\mathcal{C}, \forall t\in[t_{0}, t_{m-1})$ and the event $x(t)\in\mathcal{C}, \forall t\in[t_{0}, t_{m-1}]$ are essentially the same event.
Finally, we can recursively decompose the probability over time interval $[0, T)$:
\begin{align}
    &P\{x(t)\in\mathcal{C}, \forall t\in[0, T)\}\nonumber\\
    =&P\{x(t)\in\mathcal{C}, \forall t\in[0, t_m)\}\nonumber\\
    \ge& (1-\alpha)\cdot P\{x(t)\in\mathcal{C}, \forall t\in[t_{0}, t_{m-1})\}\nonumber\\
    \ge& (1-\alpha)\cdot(1-\alpha)\cdot P\{x(t)\in\mathcal{C}, \forall t\in[t_{0}, t_{m-2})\}\nonumber\\
    \ge& \cdots\nonumber\\
    \ge&(1-\alpha)^{m}\cdot P\{x(0)\in \mathcal{C}\}\nonumber\\
    =&(1-\alpha)^{m}
\end{align}
This completes the proof.

\end{document}